\documentclass{PoS}

\usepackage{subfig}
\usepackage{amsmath}
\usepackage{cite}

\newcommand{\be}{\begin{equation}}
\newcommand{\ee}{\end{equation}}

\title{Multi-Particle Baryon Spectroscopy }

\ShortTitle{Multi-Particle Baryon Spectroscopy }

\author{\speaker{Adrian L. Kiratidis}\\
        Special Research Centre for the Subatomic Structure of Matter,\\
        School of Chemistry and Physics, University of Adelaide, SA, 5005, Australia\\
        E-mail: \email{adrian.kiratidis@adelaide.edu.au}}
				
\author{Waseem Kamleh\\
		Special Research Centre for the Subatomic Structure of Matter,\\
        School of Chemistry and Physics, University of Adelaide, SA, 5005, Australia\\
       E-mail: \email{waseem.kamleh@adelaide.edu.au}}
       
\author{Derek B. Leinweber\\
        Special Research Centre for the Subatomic Structure of Matter,\\
        School of Chemistry and Physics, University of Adelaide, SA, 5005, Australia\\
       E-mail: \email{derek.leinweber@adelaide.edu.au}}

\author{Peter Moran}

\abstract{In Nature the excited states of the hadron spectrum appear as resonances. Consequently, there has been significant interest in studying the excited baryon spectrum using lattice QCD.  With this in mind we perform spectroscopic calculations with five-quark interpolating fields. Stochastic estimation techniques are used in order to calculate the loop propagators, with dilution in spin, colour and time implemented as a means of variance reduction. We present effective mass plots extracted from these five-quark interpolators, and examine the contributions from fully-connected and loop-containing pieces of the correlation function, keeping in mind their use in future correlation matrix studies.}

\FullConference{The 30th International Symposium on Lattice Field Theory\\
                 June 24 - 29,  2012\\
                 Cairns, Australia}

\begin{document}

\section{Introduction}
Recently the CSSM lattice collaboration has been studying the the Roper \cite{Mahbub:2010rm, Mahbub:2010jz, Mahbub:2011zza} and $\Lambda$(1405) \cite{Menadue:2011zza, Menadue:2011cw, Menadue:2011pd} resonances with conventional three-quark interpolators via correlation matrix techniques.  Although the variational techniques employed successfully isolate many states, the nearby multi-particle states are not observed.  We therefore construct five-quark operators which are expected to possess higher overlap with multi-particle states, and perform spectroscopic calculations with them with a view to including these interpolators in future variational analyses. 
We use 75 dynamical FLIC $20^3 \times 40$ lattices.  The isotropic lattice spacing is 0.126 fm at $\beta = 3.94$ and $\kappa_{u} = \kappa_{d} = 0.1324$.

\section{Lattice Techniques}
\begin{figure}[!h]
  \centering
    \includegraphics[width=0.94\textwidth, angle=0]{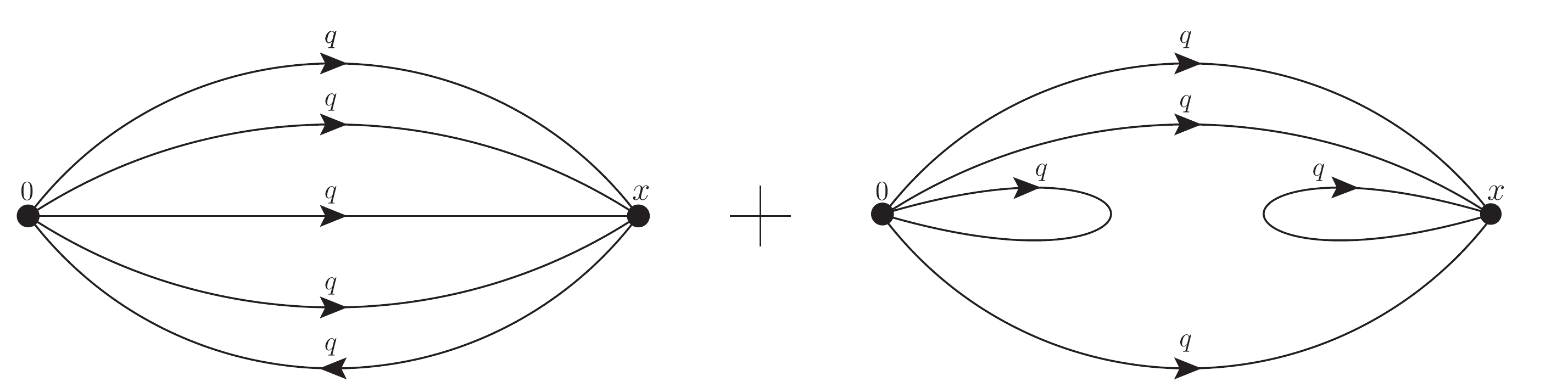}
    \caption{The ``fully-connected'' (left) and ``loop-containing'' (right) contributions to the two-point functions given.}
    \label{cfunpic}
\end{figure}

As is standard, we can derive ground state hadron masses from the parity-projected 2-point correlation function at $\vec{p} = 0$ at sufficiently large times via
\be\label{TwoptFunction}
G_{\pm}(t) = \sum_{\vec{x}}\mathrm{Tr}_{\mathrm{sp}}\big[\Gamma_{\pm}\langle\Omega\vert\chi(x)\overline{\chi}(0)\vert\Omega\rangle\big] \overset{t \rightarrow \infty}{=} \lambda_{0^\pm}\overline{\lambda}_{0^\pm}e^{-M_{0^{\pm}}t},
\ee
where $\Gamma_{\pm}$ are the parity-projection operators and $\lambda$ and $\overline{\lambda}$ are the coupling strengths that parameterise the overlap of the interpolator with a given state.  An effective mass, $M_{\pm}(t)$, can be then defined in the usual manner
\be  
M_{\pm}(t) = \ln\bigg(\frac{G_{\pm}(t)}{G_{\pm}(t+1)}\bigg).
\ee
Making use of the appropriate Clebsch-Gordan coefficients, we proceed by obtaining a five-quark Nucleon-Pion (NP)-type interpolator $\chi^{p}_{5}(x)$ (for the proton) and a five-quark Nucleon-Kaon (NK)-type interpolator $\chi^{\Lambda}_{5}(x)$ (for the $\Lambda$) with the relevant isospin.  They are given by \cite{AKThesis}
\begin{align}
\chi^{p}_{5}(x) 
&= \frac{1}{2\sqrt{3}}\,\epsilon^{abc}\,\Big\{2\big(u^{Ta}(x)C\gamma_{5}d^{b}(x)\big)d^{c}(x)\big[\bar{d}^{e}(x)\gamma_{5}u^{e}(x)\big]\nonumber\\
& \qquad - \big(u^{Ta}(x)C\gamma_{5}d^{b}(x)\big)u^{c}(x)\big[\bar{d}^{e}(x)\gamma_{5}d^{e}(x) - \bar{u}(x)^{e}\gamma_{5}u^{e}(x)\big]\Big\},
\end{align}  
and
\begin{align}\label{Lambda5QrkOpInterpolator}
\chi^{\Lambda}_{5}(x)
&= \frac{1}{\sqrt{2}}\epsilon^{abc}\Bigg[\big(u^{Ta}(x)C\gamma_{5}d^{b}(x)\big)u^{c}(x)\big[\bar{u}^{e}(x)\gamma_{5}s^{e}(x)\big]\nonumber\\
& \qquad + \big(u^{Ta}(x)C\gamma_{5}d^{b}(x)\big)d^{c}(x)\big[\bar{d}^{e}(x)\gamma_{5}s^{e}(x)\big]\Bigg].
\end{align}
One can readily see after the insertion of $\chi^{p}_{5}(x)$ or $\chi^{\Lambda}_{5}(x)$ into Eq. \ref{TwoptFunction} and the application of Wick's theorem, that the correlator has both fully-connected and loop-containing pieces as shown in Figure \ref{cfunpic}.
%
The point-to-all propagators can be calculated in the usual way by inverting the fermion matrix, the loop propagator at the source can simply be read off the point-to-all propagator, and $\gamma_{5}$-hermiticy ensures the ``backward'' propagator can be easily obtained.  However, the loop propagator at $x$ requires a source at every point on the lattice and therefore demands a different recipe.  

We therefore resort to the standard method to calculate all-to-all propagators, via stochastically estimating inverse matrix elements \cite{Foley:2005ac, Dong:1993pk}.  Our ensemble of noise vectors $\eta^{1} \ldots \eta^{N}$ is generated with $Z_{2}$ noise.  Full dilution in time, in addition to full spin and colour dilution, is performed as a means of variance reduction \cite{O'Cais:2004ww, Alexandrou:2010jr}.  That is,
\be
\eta^{a}_{\alpha}(\vec{x},t) = \sum_{b, \beta, t^{\prime}}\eta^{ab,t^{\prime}}_{\alpha\beta}(\vec{x},t).
\ee
The fermion matrix is then inverted for each of these diluted sources, obtaining the corresponding solution vectors $\chi^{cb,t^{\prime}}_{\gamma\beta}(\vec{x},t)$.  The ``dilution improved'' stochastic estimate of the loop propagator at $x$ for a single noise vector is then given by
\be\label{xxProp}
S^{ca}_{\gamma\alpha}(\vec{x}, \vec{x}) = \sum_{b, \beta, t^{\prime}}\chi^{cb,t^{\prime}}_{\gamma\beta}(\vec{x},t)\eta^{ab,t^{\prime}}_{\alpha\beta}(\vec{x},t).
\ee
%
%
\section{Results}
We proceed by presenting effective mass plots corresponding to our five-quark interpolators and the comparison to the relevant three-quark case for which we use $\chi_{1}^{p^{+}}$ and $\chi_{1}^{\Lambda^{1}}$ \cite{Leinweber:2004it}.  The $\Lambda$ is studied at the $SU(3)$ flavour limit.  We note here that although one cannot consider the fully-connected and loop-containing pieces of the five-quark correlator separately in a fully rigorous manner, the results are presented keeping in mind future correlation matrix analysis. 

\begin{figure}[H!]
\centering
\captionsetup[subfigure]{width=0.64\textwidth}
\subfloat[An effective mass plot corresponding to the five-quark proton interpolator.  The fit from the standard three-quark operator is shown in green.]
{\includegraphics[width=0.62\textwidth, angle=90]{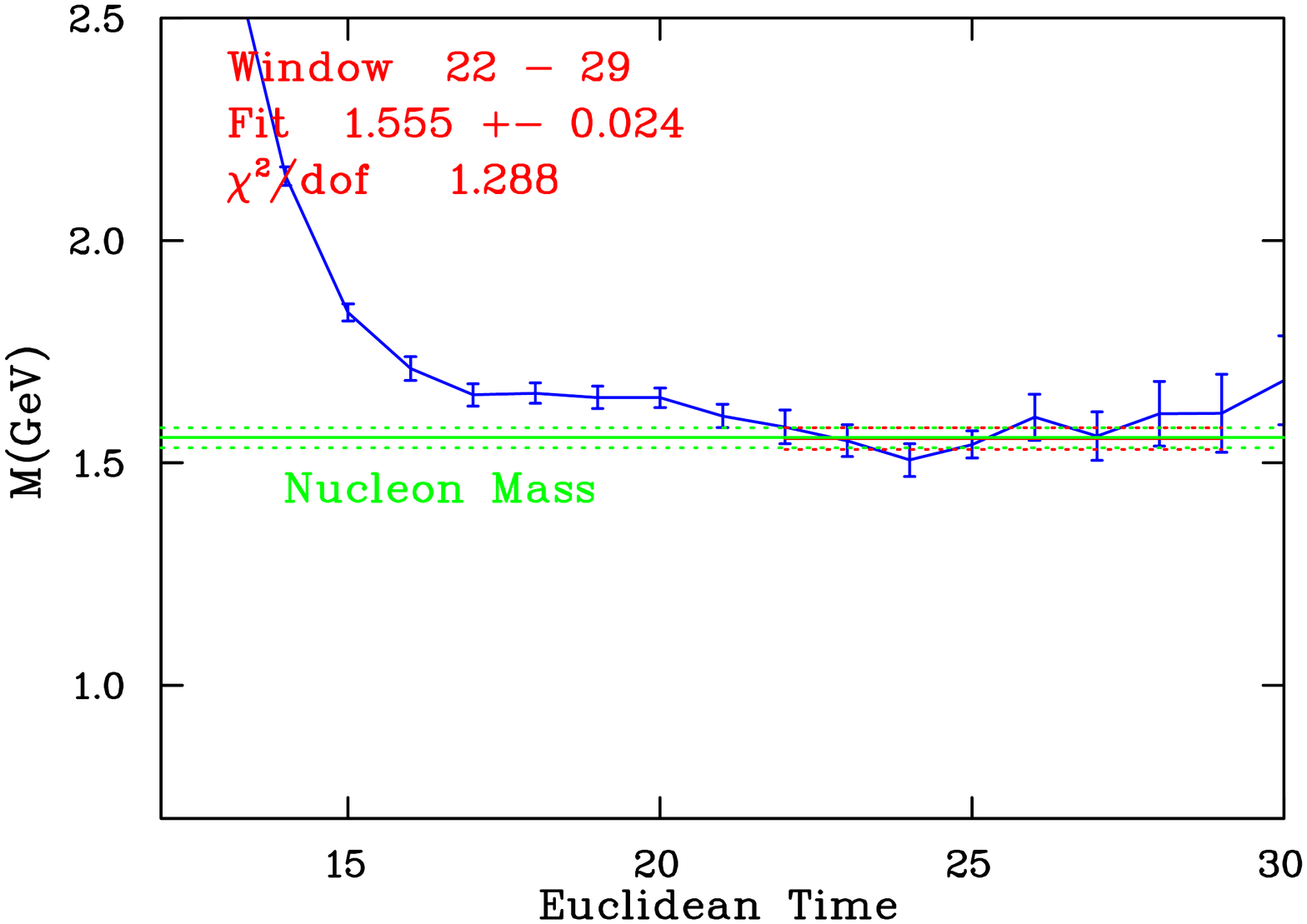}}\,\,\,\,
\subfloat[An effective mass plot showing the masses extracted from the fully-connected and loop-containing pieces of the five-quark proton correlation function.]
{\includegraphics[width=0.62\textwidth, angle=90]{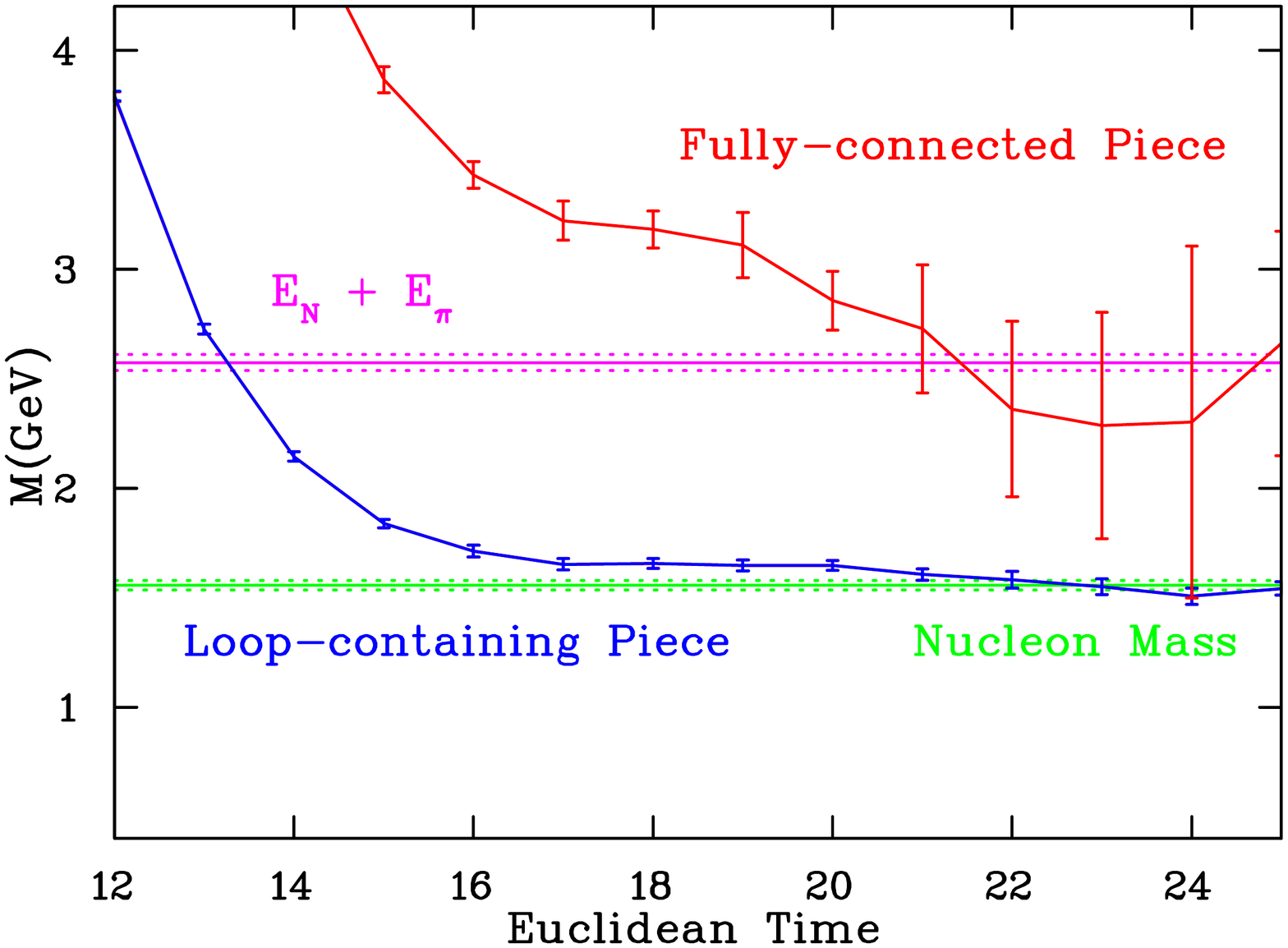}}
\caption{Effective mass plots for the proton.}
\label{ProtonPlots}
\end{figure}

\begin{figure}[H!]
\centering
\captionsetup[subfigure]{width=0.64\textwidth}
\subfloat[An effective mass plot corresponding to the five-quark $\Lambda$ interpolator.  The fit from the standard three-quark operator is shown in green.]
{\includegraphics[width=0.62\textwidth, angle=90]{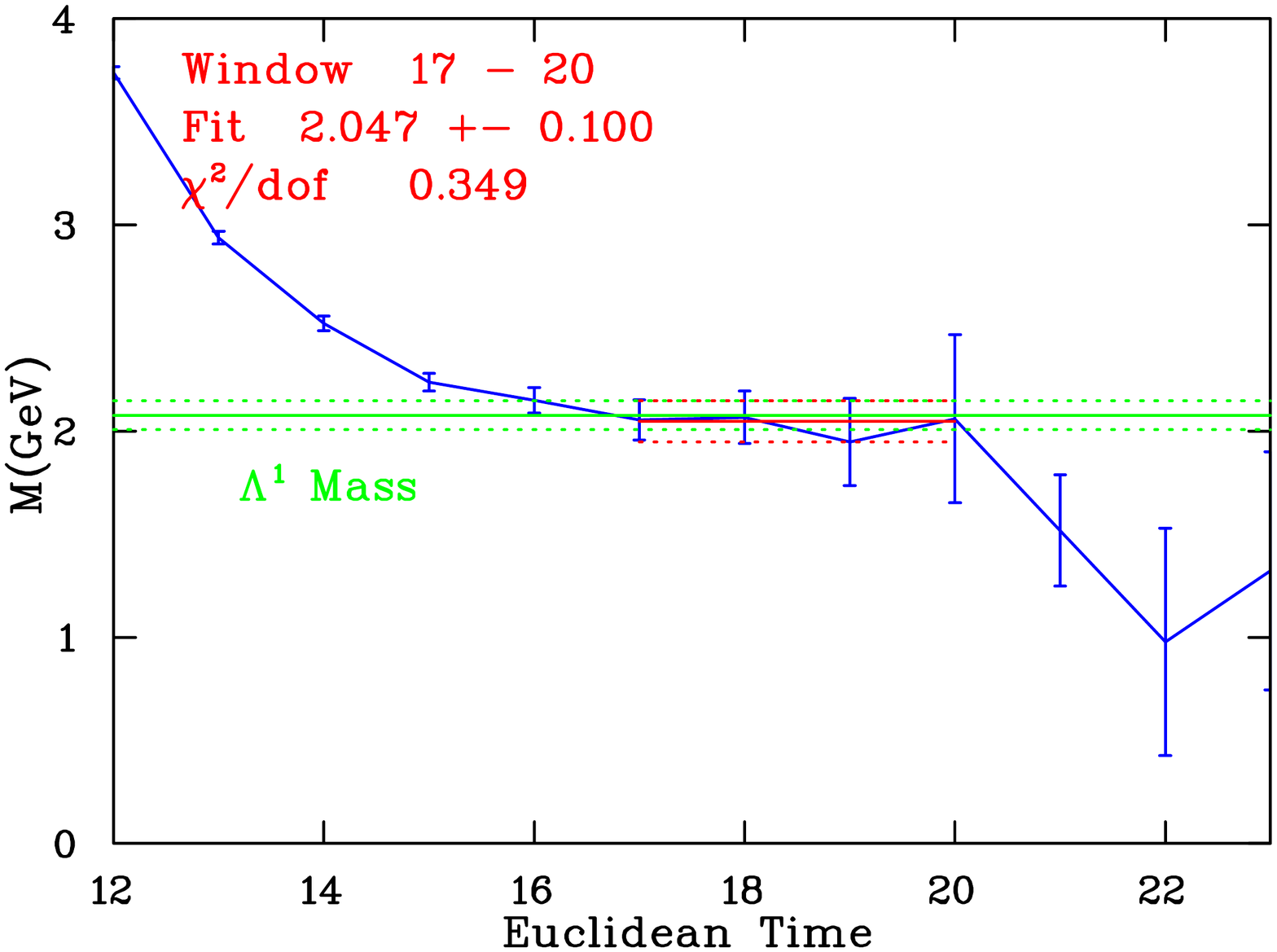}}\,\,\,\,
\subfloat[An effective mass plot showing the masses extracted from the fully-connected and loop-containing pieces of the five-quark $\Lambda$ correlation function.]
{\includegraphics[width=0.62\textwidth, angle=90]{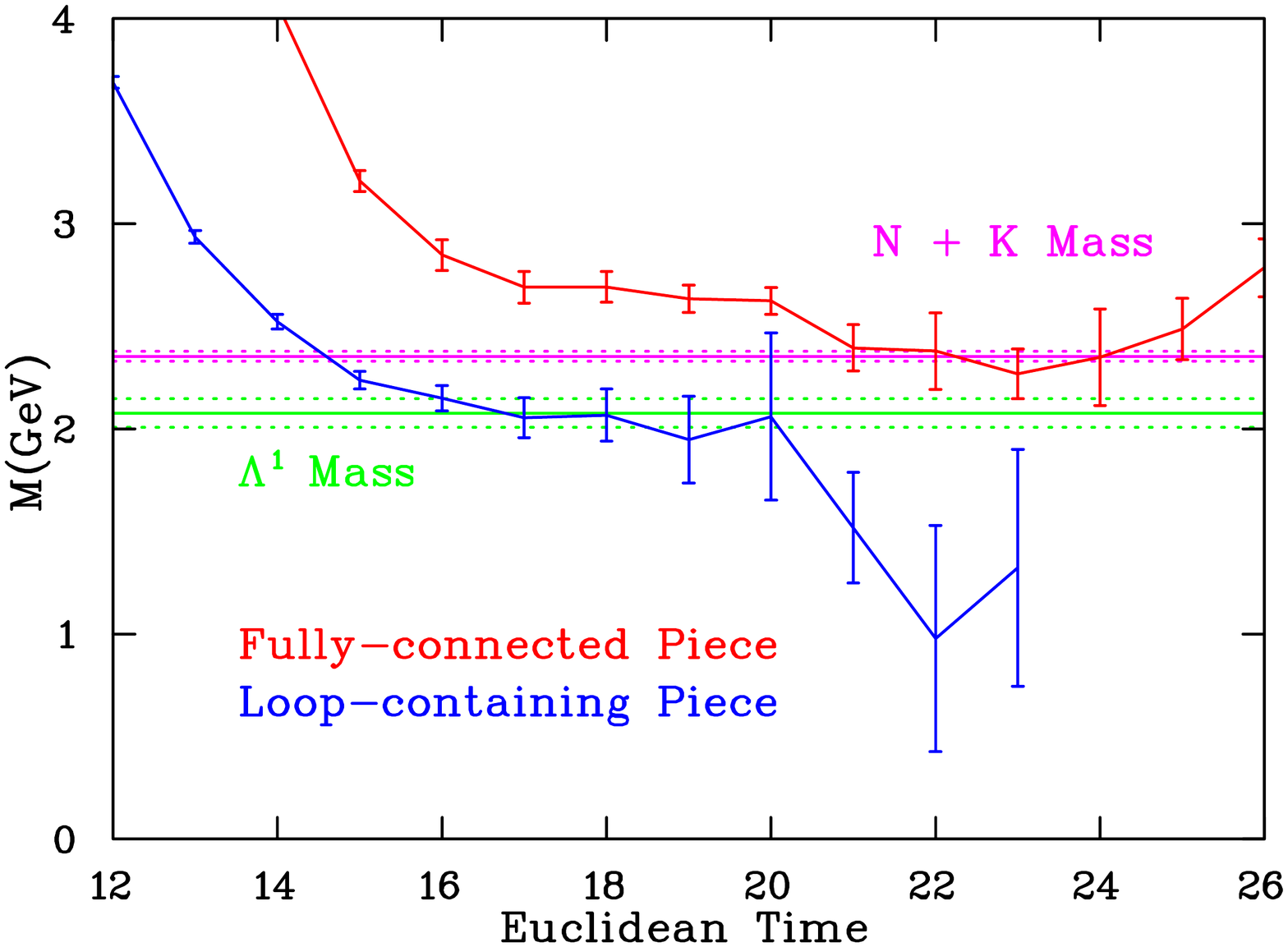}}
\caption{Effective mass plots for the $\Lambda$.}
\label{LambdaPlots}
\end{figure}

In Figure \ref{ProtonPlots} we readily observe that the five-quark mass plot is completely dominated by the loop-containing piece of the correlation function.  The mass extracted from the five-quark operator also displays good agreement with the mass obtained from the three-quark operator, indicating the possibility of quark annihilation is vital to obtaining a low-lying mass.  Furthermore, it is encouraging to note that the fully connected piece appears to be displaying substantial overlap with a more exotic state before decaying. This suggests the five-quark operator is a good candidate to facilitate the extraction of more states in future correlation matrix analyses.
%
%
The results for the $\Lambda$ shown in Figure \ref{LambdaPlots} show much the same qualitative features of the proton results in Figure \ref{ProtonPlots}.  We observe the loop-containing piece dominating, and a reproduction of the three-quark operator mass with the five-quark operator, providing explicit numerical evidence for the long-held notion that hadron spectrum does not depend on the interpolating field.

Further work in progress employs variational techniques through the use of correlation matrices \cite{Luscher:1990ck}, which will be vital in isolating multi-particle scattering states.
\newpage
\section*{Acknowledgments}
This research was undertaken on the NCI National Facility in Canberra, Australia, which is supported by the Australian Commonwealth Government.  We also acknowledge eResearch SA for grants of supercomputing time.  This research is supported by the Australian Research Council.

\end{document}